\documentclass[11pt]{article}
\renewcommand{\theequation}{\thesection.\arabic{equation}}

\usepackage{amsfonts}
\usepackage{amssymb}
\usepackage{amsbsy}
\usepackage[dvips]{graphicx}

\usepackage{dcolumn}
\newcolumntype{d}[1]{D{.}{\cdot}{#1}}

\newcommand{\BE}{\begin{equation}}
\newcommand{\EE}{\end{equation}}
\newcommand{\BA}{\begin{eqnarray}}
\newcommand{\EA}{\end{eqnarray}}
\newcommand{\half}{{\scriptstyle{\frac{1}{2}}}}

\newcommand{\msbar}{\overline{\rm MS}}
\newcommand{\smmsbar}{{\scriptscriptstyle{\overline{\rm MS}}}}
\newcommand{\kplus}{(k+1)^{th}}
\newcommand{\astar}{a^*}
\newcommand{\adot}{a^\star}
\newcommand{\ap}{a_{\rm p}}

\begin{document}

\begin{titlepage}

\vspace*{23mm}

\begin{center}
              {\huge\bf Fixed and Unfixed Points:} \\
\vspace*{2mm}
  {\LARGE\bf  Infrared limits in optimized\\
QCD perturbation theory}
\vspace{18mm}\\
{\Large P. M.  Stevenson}
\vspace{4mm}\\
{\large{\it
T.W. Bonner Laboratory, Department of Physics and Astronomy,\\
Rice University, Houston, TX 77251, USA}}

\vspace{24mm}
{\bf Abstract:}
 
\end{center}

\vspace*{1mm}
\begin{quote}
Perturbative QCD, when optimized by the principle of minimal sensitivity at 
fourth order, yields finite results for ${\cal R}_{e^+e^-}(Q)$ down to $Q=0$.  
For two massless flavours ($n_f=2$) this occurs because the couplant 
``freezes'' at a fixed-point of the optimized $\beta$ function.  
However, for larger $n_f$'s, between $6.7$ and $15.2$, the infrared limit 
arises by a novel mechanism in which the evolution of the optimized $\beta$ 
function with energy $Q$ is crucial.  The evolving $\beta$ function develops 
a minimum that, as $Q \to 0$, just touches the axis at $\ap$ (the ``pinch 
point''), while the infrared limit of the optimized couplant is at a larger 
value, $\adot$ (the ``unfixed point'').  This phenomenon results in ${\cal R}$ 
approaching its infrared limit not as a power law, but as 
${\cal R} \to {\cal R}^\star-{\rm const.}/\mid\!\ln Q \!\mid^2$.  
Implications for the phase structure of QCD as a function of $n_f$ are 
briefly considered.
\end{quote}

\end{titlepage}

\newpage

\setcounter{page}{1}

\section{Introduction}
\setcounter{equation}{0}

\begin{figure}[hbt]

\centering
\includegraphics[width=0.45 \textwidth]{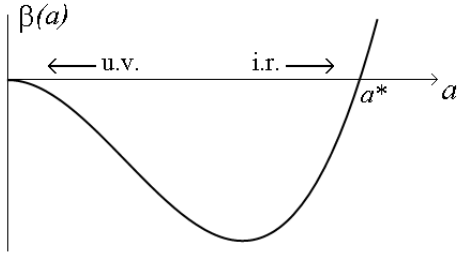}

\caption{{\textsl{ 
Conventional sketch of ``the $\beta$ function'' in an asymptotically free 
theory with an infrared fixed point.  The couplant flows to zero in 
the ultraviolet and to $\astar$ in the infrared. 
\label{betasketch}
}} }
\end{figure}

     Countless textbooks explain how key properties of a renormalizable 
field theory follow from a graph of its $\beta$ function.  
Fig.~\ref{betasketch}, for instance, supposedly represents an asymptotically 
free theory with an infrared fixed point at $a=\astar$.

The problem, though, is that there is no such thing as ``{\it the} $\beta$ 
function.''  It is a myth that there is a unique $\beta$ function  
characterizing a given theory.  In fact, away from the origin, $\beta(a)$ 
depends strongly on the arbitrary choice of renormalization scheme (RS); 
that is, it depends on the definition adopted for the renormalized coupling 
constant (couplant) $a \equiv \alpha_s/\pi$.  While the first two terms of 
$\beta(a)$ are unique, all the higher coefficients are RS dependent 
\cite{tH}.  Whether or not the $\beta$ function has a fixed point is an 
entirely RS-dependent question \cite{tH,KSS,chylafp}.

     Renormalization-group invariance \cite{GMLSP} means that any physical 
quantity, ${\cal R}$, is, in principle, independent of the RS choice.  
However, finite-order perturbative approximants to ${\cal R}$ are RS 
dependent.  The idea of ``optimized perturbation theory'' (OPT) \cite{OPT} 
is to find -- for a given ${\cal R}$ at a given energy $Q$ and at a given 
order of perturbation theory -- the ``optimal'' RS in which the perturbative 
approximant is locally invariant; {\it i.e.}, stationary under small changes 
of RS.  At second (next-to-leading) order this optimization is simply a 
precise formulation of the familiar and powerful idea that the renormalization 
scale $\mu$ should not be kept fixed but should ``run'' with the experimental 
energy scale $Q$.  At higher orders, though, the optimization procedure also 
determines optimal values for the higher-order $\beta$-function coefficients, 
and these evolve as the energy $Q$ is changed.  Thus, the optimized $\beta$ 
function itself evolves.  

\begin{figure}[hbt]


\centering
\includegraphics[width=0.5 \textwidth]{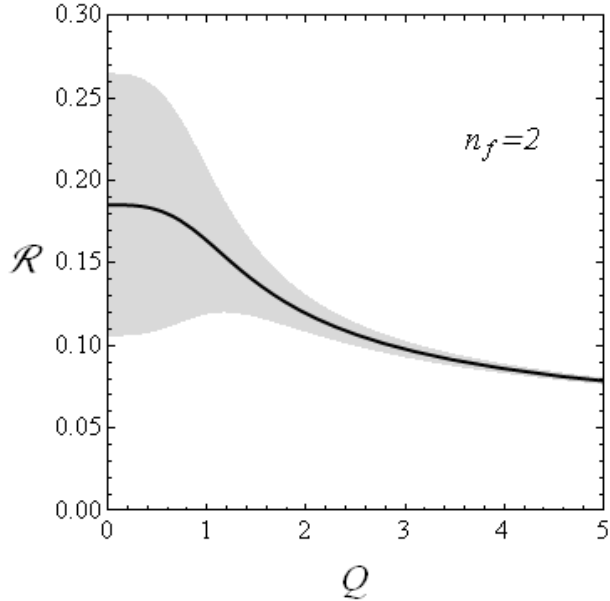}
\caption{{\textsl{OPT results for ${\cal R}_{e^+e^-}$(non-singlet) for 
$n_f=2$.  The energy $Q$ is in units of $\tilde{\Lambda}_{{\cal R}}$, 
see Eq.~(\ref{Lameff}).  The shaded region indicates the error estimate. 
\label{RvsQnf2}
}} }
\end{figure}

     Hitherto this last point had seemed -- even to this author -- a 
technicality, unlikely to overthrow the basic picture that a finite infrared 
limit in QCD only occurs if ``the $\beta$ function'' has a fixed point.  
Such a fixed-point limit of OPT was analyzed in Ref.~\cite{KSS} and was 
later found to occur in QCD in the third-order $R_{e^+e^-}$ case 
\cite{CKL,lowen}.  The recent calculation of the fourth-order correction 
to $R_{e^+e^-}$ \cite{r3calc} has allowed us in Ref.~\cite{OPTnew} to 
investigate OPT at fourth order.  For the phenomenologically relevant case 
of two massless flavours ($n_f=2$) we again found fixed-point behaviour with 
the couplant freezing to a modest value, with the third-order result 
$0.3 \pm 0.3$ \cite{lowen} now refined to $0.2 \pm 0.1$ \cite{OPTnew}.  
See Fig.~\ref{RvsQnf2}.

\begin{figure}[!hbt]


\centering
\includegraphics[width=0.5 \textwidth]{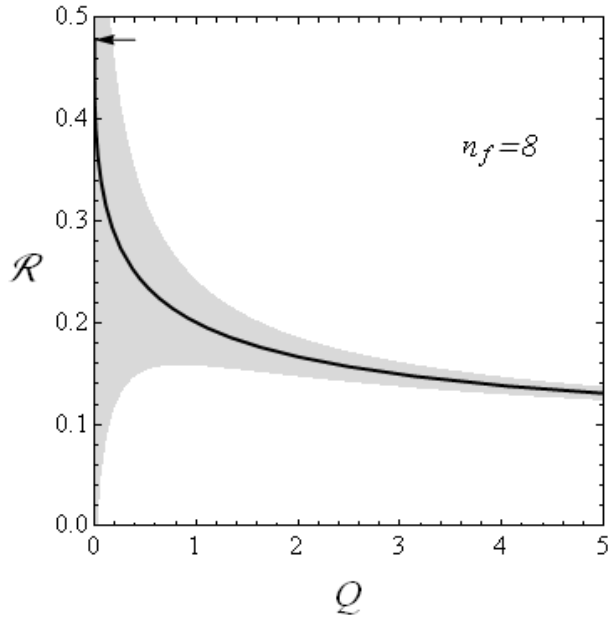}
\caption{{\textsl{As Fig.~\ref{RvsQnf2} but for $n_f=8$.  The arrow 
indicates the infrared limit. 
\label{RvsQnf8}
}} }
\end{figure}

   Continuing our investigation to higher $n_f$ values, however, produced 
a surprise: a finite infrared limit in OPT can also occur by a quite 
different mechanism in which the evolution of the $\beta$ function 
plays an essential role.  This ``pinch mechanism'' produces an extreme 
``spiking,'' rather than a ``freezing,'' of the couplant as $Q \to 0$; 
see Fig.~\ref{RvsQnf8}.  The main purpose of this paper is to describe the 
pinch mechanism and to present numerical results for the infrared limit 
as a function of $n_f$.

     In discussing the infrared behaviour of perturbation theory in QCD, one 
must of course recognize that the results are not directly physical.  There 
exist large nonperturbative, higher-twist terms that perturbation theory 
is completely blind to.  However, it is a longstanding idea \cite{bloom,PQW} 
that perturbation theory corresponds to some kind of average over hadronic 
resonances.  As shown in Ref.~\cite{lowen}, the low-energy $e^+e^-$ data 
agrees very nicely, in this sense, with the prediction of OPT that the 
couplant freezes to a modest value.  Moreover, there is a wealth of 
phenomenological evidence for freezing.\footnote{
Recently it was pointed out that a fixed point in the $n_f=3$ theory provides 
a simple and appealing explanation of the $\Delta I = \frac{1}{2}$ rule for 
Kaon decays \cite{Crewther}.}
We therefore believe that there is some real-world 
relevance to studying the infrared behaviour of perturbation theory.  

     It is also interesting theoretically to consider QCD with $n_f$ 
flavours of massless quarks for various $n_f$.  One reason is to compare 
with extrapolations from $n_f=16\half$ (the Banks-Zaks (BZ) or ``small-$b$'' 
expansion \cite{BZ}), or from $n_f=-\infty$ (the ``large-$b$'' approximation
\cite{largeb,LTM}).  The other reason is the whole issue of the phase 
structure of QCD, and other gauge theories, as a function of $n_f$ 
\cite{Appelquist} -- \cite{MC}.

     We begin by briefly reviewing the key ingredients of OPT \cite{OPT} 
in Sect.~2.  (See Ref.~\cite{OPTnew} for a fuller account.) Sections 3 and 4 
respectively describe the fixed-point and pinch mechanisms whereby 
fourth-order OPT produces a finite $Q=0$ limit.  Numerical results are 
presented in Sect.~5.  Sect.~6 describes the approach to the $Q=0$ limit 
and Sect.~7 briefly discusses the possible implications of the results.

\section{Optimized perturbation theory}
\setcounter{equation}{0}

  The $\beta$ function, of some general RS, is written as
\BE
\beta(a) \equiv \mu \frac{da}{d\mu} = - ba^2 B(a),
\EE
with
\BE
B(a)=(1+ca + c_2a^2+c_3a^3+\ldots).
\EE
The first two coefficients of the $\beta$ function are RS invariant 
\cite{tH} and are given by \cite{bcalc,ccalc}
\BE
b = \frac{(33-2n_f)}{6}, \quad\quad c= \frac{153 - 19 n_f}{2(33-2n_f)}.
\EE
The higher $\beta$-function coefficients $c_2, c_3, \ldots$ are RS dependent:
together with $\tau$, 
\BE
\tau \equiv  b \ln(\mu/{\tilde{\Lambda}}),
\EE
they parametrize the RS choice. 

     The $\tilde{\Lambda}$ parameter in $\tau$ arises as the constant of 
integration in the integrated $\beta$-function (int-$\beta$) equation
\BE
\label{intbeta}
\tau = K(a),
\EE
where 
\BE
\label{K}
K(a) \equiv \frac{1}{a} + c \ln (\mid\!c\!\mid\! a) - 
\int_0^a \frac{dx}{x^2} \left( \frac{1}{B(x)}-1 + c x \right).
\EE
This form of $K(a)$, completely equivalent to our previous definition 
\cite{OPT,OPTnew}, is more convenient when $c$ may be negative.  
The $\tilde{\Lambda}$ parameter thus defined 
%
%
is RS dependent, but it can be converted between different schemes 
{\it exactly} by the Celmaster-Gonsalves relation~\cite{CG}. 

     The physical quantity considered here is the $e^+e^-$ hadronic 
cross-section ratio $R_{e^+e^-}(Q)$ at a total c.m. energy $Q$.  Neglecting 
quark masses, this has the form $R_{e^+e^-} = 3 \sum q_i^2 (1 + {\cal R})$ 
with
\BE 
{\cal R} = a(1+r_1 a+ r_2 a^2 + r_3 a^3 + \ldots ).
\EE
(Actually, in this paper we will consider only the ``non-singlet'' part of 
${\cal R}$; that is, we drop the terms proportional to 
$\left( \sum q_i \right)^2$.  This makes very little difference for 
$0 \le n_f \le 6$ and allows us to discuss higher $n_f$'s without needing to 
specify the electric charges of the fictitious, additional quarks.)  

     Since it is a physical quantity, ${\cal R}$ satisfies a set of RG 
equations~\cite{OPT}
\BA
\label{rga}
 \frac{\partial {\cal R}}{\partial \tau} =
\left( \left. \frac{\partial}{\partial \tau} \right|_a + 
\frac{\beta(a)}{b} \frac{\partial}{\partial a} \right) {\cal R} \, = 0, 
 & \quad\quad ``j=1{\mbox{\rm''}}, & \nonumber \\ 
 & & \\
\frac{\partial {\cal R}}{\partial c_j} =
\left( \left. \frac{\partial}{\partial c_j} \right|_a + 
\beta_j(a) \frac{\partial}{\partial a} \right) {\cal R} = 0, & \quad\quad 
j = 2,3,\ldots. & \nonumber
\EA
The first of these (``$j=1$'') is the familiar RG equation expressing the 
invariance of ${\cal R}$ under changes of renormalization scale $\mu$.  
The other equations express the invariance of ${\cal R}$ under other changes 
in the choice of RS.  The $\beta_j(a)$ functions, defined as 
$\partial a/\partial c_j$, are given by
\BE
\beta_j(a) \equiv \frac{a^{j+1}}{(j-1)} B_j(a),
\EE
with
\BE
\label{Bj}
B_j(a) = \frac{(j-1)}{a^{j-1}} B(a) I_j(a),
\EE
where 
\BE
\label{Ij}
I_j(a) \equiv \int_0^a dx \, \frac{x^{j-2}}{B(x)^2}.
\EE
The $B_j(a)$ functions have expansions that start $1+O(a)$.  (Note that 
for $j\to1_+$ one naturally finds $B_1(a)=B(a)$.)

      The RG equations (\ref{rga}) imply that certain combinations of 
${\cal R}$ and $\beta$-function coefficients are RS invariants.  
Up to fourth order these are:
\BA
\tilde{\rho}_1 & = & c, \quad \quad {\mbox{\rm and}} \quad\quad 
\boldsymbol{\rho}_1(Q)= \tau-r_1 ,
\nonumber \\
\tilde{\rho}_2 & = & c_2+r_2-c r_1-r_1^2, 
\label{rhodefs}
\\
\tilde{\rho}_3 & = & c_3 + 2 r_3 -2 c_2 r_1 -6 r_2 r_1 + c r_1^2 + 4 r_1^3 .
\nonumber 
\EA
The numerical values of the $\tilde{\rho}_2, \tilde{\rho}_3$ invariants 
(see Table~\ref{invtable} below) can be obtained from the $\msbar$ 
calculations of $c_2,c_3$ \cite{c2calc,c3calc} and $r_1,r_2,r_3$ 
\cite{r1calc,r2calc,r3calc}, with all dependence on the arbitrary $\msbar$ 
choice dropping out.   

      $Q$ dependence enters only through $\boldsymbol{\rho}_1(Q)$, 
which can be conveniently expressed as 
\BE
\label{Lameff}
\boldsymbol{\rho}_1(Q) = b \ln (Q/\tilde{\Lambda}_{{\cal R}}),
\EE
where $\tilde{\Lambda}_{{\cal R}}$ is a characteristic scale specific to 
the particular physical quantity ${\cal R}$.  It can be related back to the 
traditionally defined $\Lambda_\smmsbar$ parameter by the exact relation 
\BE
\ln (\tilde{\Lambda}_{{\cal R}}/\Lambda_\smmsbar) = 
\frac{r_1^{\smmsbar}}{b} - (c/b) \ln (2\!\mid\!c\!\mid\!/b).
\EE
Note that the infrared limit corresponds to 
$\boldsymbol{\rho}_1(Q) \to -\infty$.

   The $\kplus$-order approximation, ${\cal R}^{(k+1)}$, in some general RS, 
is defined by truncating the ${\cal R}$ and $\beta$ series after the $r_k$ and 
$c_k$ terms, respectively.  Because of these truncations, the resulting 
approximant depends on RS\@.  ``Optimization'' \cite{OPT} corresponds to 
finding the stationary point where the approximant is locally insensitive 
to small RS changes, i.e., finding the ``optimal'' RS in which the RG 
equations (\ref{rga}) are satisfied by ${\cal R}^{(k+1)}$ with no remainder. 
The resulting optimization equations \cite{OPT} were recently solved for  
the optimized $\bar{r}_m$ coefficients \cite{OPTnew}.

   [A note about notation:  An overbar can be used to explicitly distinguish 
optimized from generic quantities, but we shall generally omit these below, 
leaving it understood that all quantities are the optimized ones at 
$\kplus$ order.  The one exception is the symbol ``$a$,'' which we employ 
merely as a dummy argument.  Thus, we can discuss ``the $\beta(a)$ function'' 
in basically the traditional sense as a function of a single variable, $a$, 
with definite coefficients, $c_j=\bar{c}_j$; the key difference, though, is 
that the $\bar{c}_j$ coefficients will themselves evolve as the energy $Q$ 
changes.]

   The optimized $r_m$ coefficients are given in terms of the optimized 
couplant $\bar{a}$ and the optimized $c_j$ coefficients by \cite{OPTnew}:  
\BE 
\label{formula}
(m+1)r_m \bar{a}^m = 
\frac{1}{B_k(\bar{a})} \left( H_{k-m}(\bar{a}) - H_{k-m+1}(\bar{a}) \right), 
\quad\quad\quad m=0,1,\ldots,k,
\EE 
where 
\BE
\label{Hdef}
H_i(a) \equiv \sum_{j=0}^{k-i} c_j a^j \left( \frac{i-j-1}{i+j-1} \right) 
B_{i+j}(a), 
\quad\quad\quad i=(1),2,\ldots,k,
\EE
with $c_0 \equiv 1$, $c_1\equiv c$.  $H_1$ is to be understood as the limit 
$i \to 1$ of the above formula, and $H_0 \equiv 1$ and $H_{k+1} \equiv 0$.  
At fourth order ($k=3$) the $H$'s are explicitly given by 
\BE
\label{Hdefk3}
H_1 = B - c a B_2-c_2 a^2 B_3, \quad H_2=B_2, \quad 
H_3=B_3, 
\EE
and the optimized $r_m$ coefficients are given by
\BE
2 r_1 a   =\frac{H_2-H_3}{H_3}, \quad
3 r_2 a^2 =\frac{H_1-H_2}{H_3}, \quad
4 r_3 a^3 =\frac{1-H_1}{H_3},
\EE
with $a=\bar{a}$.  

   The optimized $r_m$ and $c_j$ coefficients must also be constrained to 
yield the $\tilde{\rho}_n$ invariants of Eq.~(\ref{rhodefs}).  The iterative 
algorithm outlined in Ref.~\cite{OPTnew} can be used to solve numerically 
for the optimized coefficients, and thereby obtain the optimized result, at 
any given $Q$ value.  In the $Q \to 0$ limit these steps can be carried out 
analytically, as discussed in the next two sections.

\section{Fixed-point mechanism}
\setcounter{equation}{0}


\begin{figure}[htb]

\centering
\includegraphics[width=0.64 \textwidth]{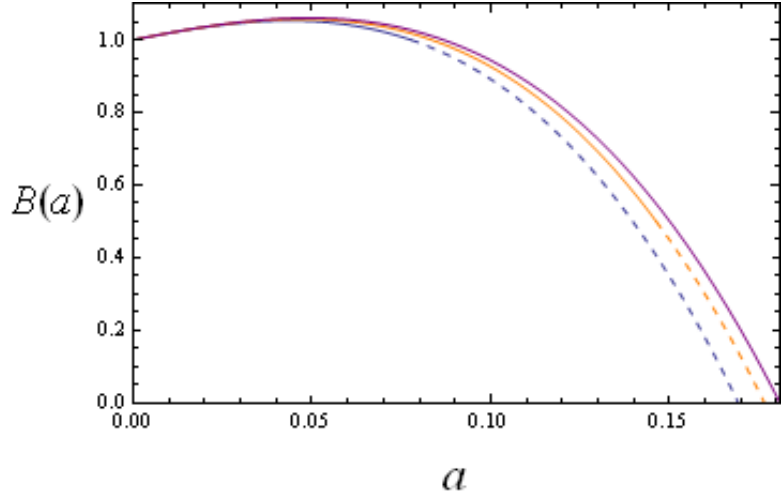}

\caption{{\textsl{
The evolving optimized $B(a) \equiv \beta(a)/(-b a^2)$ function at fourth 
order for $n_f=2$.  The upper, solid curve is the $Q=0$ limiting form with 
a fixed point at $\astar=0.180844$.  The two lower curves correspond to 
larger $Q$ values, and are shown dashed when $a > \bar{a}$. 
\label{Bfignf2} }}}
\end{figure}

   A finite $Q \to 0$ limit for ${\cal R}(Q)$ can occur by essentially 
the familiar fixed-point mechanism, with the optimized $B(a)$ function 
manifesting a simple zero at $a=\astar$ (see Fig.~\ref{Bfignf2}).  
The limiting behaviour can be analyzed as follows \cite{KSS}.  
For $a$ close to $\astar$ one can linearize $B(a)$ as
\BE
B(a) \approx \sigma (\astar - a),
\EE
where $\sigma$ is some positive constant (directly related to 
$\gamma^*$, the slope of the $\beta$ function at its fixed point; 
$\gamma^*=b {\astar}^2 \sigma$).  The integrals $I_j(a)$ of 
Eq.~(\ref{Ij}) will then diverge in the infrared limit, $a \to \astar$:
\BE
I_j(a) \to \int_0^a \! dx \frac{x^{j-2}}{\sigma^2(\astar-x)^2} 
\to  \frac{{\astar}^{j-2}}{\sigma^2} \frac{1}{(\astar-a)}.
\EE
Substituting in $B_j(a)$, Eq.~(\ref{Bj}), one finds that the 
$\frac{1}{(\astar-a)}$ factor is cancelled by the $(\astar-a)$ factor in 
$B(a)$, yielding
\BE
\label{Bjfp}
B_j(a) \to \frac{(j-1)}{\sigma \astar}.
\EE
This result corresponds to 
$\partial \astar / \partial c_j \to {\astar}^j/\sigma$, which indeed follows 
directly \cite{KSS} by taking $\partial / \partial c_j$ (with the other 
$c_i$'s held constant) of the $\kplus$-order fixed-point condition 
\BE
\label{fpeq1}
B(\astar)=\sum_{i=0}^k c_i {\astar}^i=0.
\EE
The slope parameter $\sigma$ is given by 
\BE
\label{sigma}
\sigma= \left. -B'(a)\right|_{a=\astar} = - \sum_{j=0}^k j c_j {\astar}^{j-1} 
= \sum_{j=0}^{k-1} (k-j) c_j {\astar}^{j-1},
\EE
where the last step uses the fixed-point condition (\ref{fpeq1}) to 
eliminate $c_k$.  With the limiting $B_j$'s from Eq.~(\ref{Bjfp}) one can 
construct the $H_j$'s and hence the limiting values of the optimized $r_m$ 
coefficients. 

    At fourth order ($k=3$) one obtains
\BE
r_1^* = - \frac{1}{4 \astar}, \quad 
r_2^* = - \frac{(1+ c \astar + 2 c_2^* {\astar}^2)}{6 {\astar}^2}, \quad
r_3^* = - \frac{3}{8} c_3^*.
\EE
By substituting in the definitions of $\tilde{\rho}_2, \tilde{\rho}_3$, 
Eq.~(\ref{rhodefs}), one can then find $c_2^*, c_3^*$ in terms of $\astar$ and 
those invariants.  The fixed-point condition above can then be expressed 
entirely in terms of invariants as \cite{KSS}
\BE 
\label{astar3}
\frac{83}{64} + \frac{13}{16} c a^* + \frac{3}{4} \tilde{\rho}_2 {a^*}^2 
+ 2 \tilde{\rho}_3 {a^*}^3 =0.
\EE
The relevant $\astar$ is the smallest positive root of this equation.
The final result for the limiting value of ${\cal R}$ at fourth order 
can then be simplified to \cite{KSS}
\BE
\label{Rstar3}
{{\cal R}^*} = a^* \left( \frac{249}{256} + \frac{13}{64} c a^* + 
\frac{1}{16} \tilde{\rho}_2 {a^*}^2 \right).
\EE
Eq.~(\ref{astar3}) turns out to have no acceptable root when 
$n_f$ is $7,\ldots,15$.  (For $n_f=15$ there is a positive root but it 
gives a negative slope $\sigma$, which is unphysical.)  Nevertheless, 
going to ever lower $Q$ values with the optimization procedure, one does find 
that the optimized result remains bounded as $Q \to 0$.  How this happens 
is the topic of the next section.

\section{Pinch mechanism}
\setcounter{equation}{0}

\begin{figure}[!hbt]

\centering
\includegraphics[width=0.64 \textwidth]{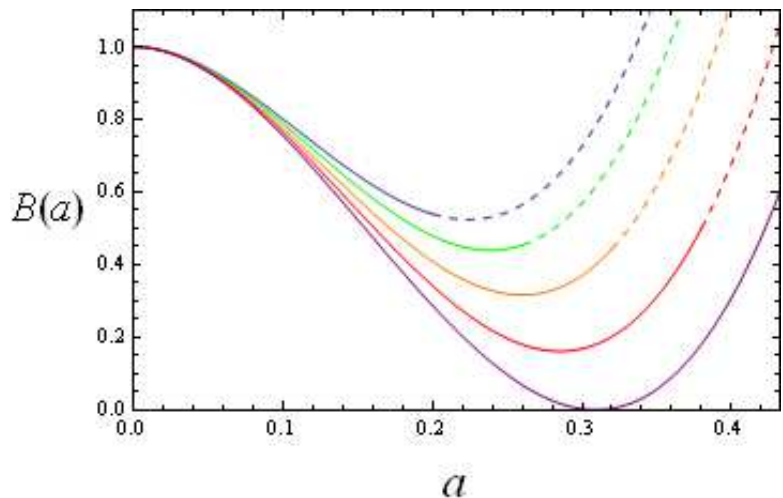}
\caption{{\textsl{
The evolving optimized $B(a)\equiv \beta(a)/(-b a^2)$ function at fourth order 
for $n_f = 8$.  The curves, from top to bottom, are for descending $Q$ values. 
They are shown dashed for $a>\bar {a}$.  The lowest curve is the 
infrared-limiting form, with the pinch point at $\ap = 0.3094$ and the 
unfixed point at $\adot = 0.432267$.
\label{Bfignf8} }}}
\end{figure}

    The essence of the pinch mechanism is illustrated in Fig.~\ref{Bfignf8}, 
which shows the evolution of the optimized $B(a)$ function in the $n_f=8$ 
case.  As $Q$ is lowered the optimized $c_2, c_3$ coefficients change so that 
$B(a)$ develops a minimum --- which, in the limit $Q \to 0$, just 
touches the horizontal axis at a ``pinch point,'' $\ap$.  Although this point 
is then a double root of $B(a)=0$, it does {\it not} represent a fixed point.  
The infrared-limit of the optimized couplant is not $\ap$ but a larger value, 
$\adot$, dubbed the ``unfixed point'' to stress that it is not a zero of the 
$\beta$ function.

\vspace*{6mm}


\newpage

      One can understand this infrared behaviour analytically as follows.  
$B(a)$ can be approximated around its minimum (at, or nearly at, the pinch 
point $\ap$) by 
\BE
\label{Bform}
B(a) \approx \eta \left( (a-\ap)^2 + \delta^2 \right),
\EE
where $\delta \to 0$ as $Q \to 0$ and $\eta$ is some positive constant.  
Thus the integral for the $K(a)$ function in Eq.~(\ref{K}) becomes dominated 
by a ``resonant peak'':
\BE 
\label{pinchdel}
-\int \frac{dx}{x^2} \frac{1}{\eta \left( (x-\ap)^2 + \delta^2 \right)} 
\approx -\frac{1}{\ap^2 \eta} \frac{\pi}{\delta} + \mbox{\rm finite}.
\EE
Therefore, in the $Q \to 0$ limit (where $\boldsymbol{\rho}_1(Q) =K(a)-r_1$ 
tends to $-\infty$), the $\delta$ parameter vanishes  
$ \propto 1/\!\mid\! \ln Q \!\mid$.  

     The integrals $I_j(a)$ of Eq.~(\ref{Ij}) are also dominated by a 
huge peak in their integrands around $\ap$:
\BE
\label{Ijform}
I_j(a) \approx \int \! dx 
\frac{x^{j-2}}{\left( \eta \left( (a-\ap)^2 + \delta^2 \right) \right)^2} 
\approx \frac{\ap^{j-2}}{\eta^2} \frac{\pi}{2 \delta^3}.
\EE
One can thus obtain the $\delta \to 0$ behaviour of the $B_j(a)$ and 
hence the $H_j$ functions.  (Note that the $B(a)/a^{j-1}$ factor in 
Eq.~(\ref{Bj}) will involve the limiting value of $a$, which is $\adot$ and 
not $\ap$.)  While the $B_j$'s and $H_j$'s diverge, the $1/\delta^3$ 
factors cancel out, as does $\eta$, in Eq.~(\ref{formula}), leaving finite 
limiting values for the optimized $r_m$ coefficients.  

    At fourth order ($k=3$) one finds
\BA
2 r_1^\star & = & \frac{(\adot-2\ap)}{2 \adot \ap}, \nonumber \\
\label{rlim}
3 r_2^\star & = & -\frac{1}{2 \adot \ap} 
\left( 1+ c \adot + 2 c_2^\star \adot \ap \right), \\
4 r_3^\star & = & \frac{(c+ 2 c_2^\star \ap)}{2 \adot \ap}. \nonumber
\EA
The infrared limit of the fourth-order $B(a)$ function is 
\BE
B^\star(a) = 1+ c a + c_2^\star a^2 + c_3^\star a^3.
\EE
The pinch point $\ap$ is where this function touches the $a$ axis (see 
Fig.~\ref{Bfignf8}) and hence satisfies the two equations $B^\star(a)=0$ 
and $dB^\star/da=0$ at $a=\ap$.  These two equations yield 
\BA 
c_2^\star & = & - \frac{(3+2 c \ap)}{\ap^2}, \nonumber \\
\label{clim} \\
c_3^\star & = & \frac{(2+c \ap)}{\ap^3}. \nonumber
\EA
Substituting Eqs.~(\ref{rlim}) and (\ref{clim}) into the definitions of the 
$\tilde{\rho}_2$ and $\tilde{\rho}_3$ in Eq.~(\ref{rhodefs}) yields two 
equations:
\BE
12 \ap^2 -4 (1+ 6 c \ap) \adot \ap 
+ (99 + 84 c \ap + 48 \tilde{\rho}_2 \ap^2 ) {\adot}^2 =0,
\EE
\BE
8 \ap^3 -4(1+ c \ap) \adot \ap^2 + 2(13+12 c \ap) {\adot}^2 \ap
+ (-33-21 c \ap + 16 \tilde{\rho}_3 \ap^3 ) {\adot}^3 =0.
\EE
These two equations determine $\ap$ and $\adot$ in terms of the invariants
$c, \tilde{\rho}_2, \tilde{\rho}_3$.  One can manipulate these equations 
to find $\ap$ in terms of $\adot$ as
\BE
\ap = \frac{\adot (62+14 c \adot + (21 c^2-148 \tilde{\rho}_2) {\adot}^2 
-6(7 \tilde{\rho}_2 c - 33 \tilde{\rho}_3) {\adot}^3 )}
{2(2-31 c \adot + 3 (4 \tilde{\rho}_2 - c^2) {\adot}^2 
+ 4(6 \tilde{\rho}_2 c + \tilde{\rho}_3) {\adot}^3 
+ 12 (4 \tilde{\rho}_2^2 - 7 \tilde{\rho}_3 c) {\adot}^4 )},
\EE
with $\adot$ given by a 6th-order polynomial equation:
\BA
0 & = & 
11680 + 2224 c \adot + 3(5997 c^2 -17264 \tilde{\rho}_2) {\adot}^2 
+ \nonumber \\
& & 
{} + 2(8235 c^3 - 33624 \tilde{\rho}_2 c + 36976 \tilde{\rho}_3) {\adot}^3 
+ \nonumber \\
& & 
{} + 18 (147 c^4-2184 \tilde{\rho}_2 c^2 + 4640 \tilde{\rho}_2^2 
+ 502 \tilde{\rho}_3 c) {\adot}^4 
+  \\
& & 
{} + 324 (-49 \tilde{\rho}_2 c^3 +152 \tilde{\rho}_2^2 c 
+161 \tilde{\rho}_3 c^2 - 528 \tilde{\rho}_3 \tilde{\rho}_2) {\adot}^5 
+ \nonumber \\
& & 
{} + 108(-147 \tilde{\rho}_2^2 c^2 + 528 \tilde{\rho}_2^3 
+ 343 \tilde{\rho}_3 c^3 -1386 \tilde{\rho}_3 \tilde{\rho}_2 c 
+ 1089 \tilde{\rho}_3^2) {\adot}^6 \nonumber .
\EA 
The final result for the infrared limit of ${\cal R}$ at fourth order 
can be expressed as 
\BE
{\cal R}^\star = \frac{\adot(2 \adot \ap + 12 {\ap}^2 + 
3 {\adot}^2 (2+c \ap))}{24 {\ap}^2}.
\EE
Note that $\adot \ge \ap$ is needed for this solution to be relevant.  One 
can check that the special case $\adot=\ap$ is indeed the boundary between 
the pinch mechanism and the fixed-point mechanism, and corresponds to 
where $\gamma^*=0$.  From such an analysis one can determine the precise 
$n_f$ values where the switchover from one mechanism to the other takes place.

     It is possible, in principle, for the pinch mechanism to occur 
at third order; see Appendix A.

\section{Numerical results}
\setcounter{equation}{0}

    The inputs to our numerical calculations are collected in 
Table~\ref{invtable}, which lists the RS-invariant quantities 
$c$, $\tilde{\rho}_2$, $\tilde{\rho}_3$ for integer $n_f$ from $0$ to $16$.
These values are obtained from the Feynman-diagram calculations of 
Ref.~\cite{r3calc} and earlier authors 
\cite{ccalc}, \cite{c2calc}--\cite{r2calc}.  
(The singlet terms, proportional to $(\sum q_i)^2$ have been dropped.)

    Table~\ref{IRtable} gives our results for the infrared limit of 
${\cal R}$ for $n_f=0,\ldots,16$.  The quoted error estimate on 
${\cal R}$ corresponds to the last term, $r_3 a^4$, of the truncated 
perturbation series, evaluated in the optimized RS \cite{lowen,OPTnew}.  
Also listed are values of the fixed-point, or the unfixed-point and 
pinch-point.  (The critical exponent $\gamma^*$ will be discussed later.)  
The fixed-point mechanism operates for $n_f < 6.727$, then the pinch 
mechanism takes over until $n_f=15.191$, when the fixed-point mechanism 
returns and operates until $n_f=16\half$ when $\astar \to 0$.    

     Table~\ref{coeffstable} gives the optimized coefficients, weighted by 
the appropriate power of $\bar{a}$, in both the $\beta$-function and 
${\cal R}$ series.  This information is important for anyone wishing to 
check our results and also displays the behaviour of the truncated series 
for both ${\cal R}$ and $B(a)$.  This behaviour is, at best, only marginally 
satisfactory: Clearly, by going to the $Q \to 0$ limit we are pushing 
low-order perturbation theory well beyond its comfort zone.  
Nevertheless, all things considered, we believe that the results are 
credible within the large uncertainties quoted in Table~\ref{IRtable} and 
illustrated in Figs.~\ref{RvsQnf2} and \ref{RvsQnf8}.  In particular, we 
believe that the dramatic $Q \to 0$ spike 
produced by the pinch mechanism is real;  the very large error estimate 
just cautions that the height of the spike is very uncertain; it might 
be somewhat smaller, or it might well be considerably bigger.    

\vspace*{4mm}
   
\begin{table}[htbp]
\begin{center}
\begin{tabular}[b]{|r||d{6}|d{6}|d{5}|}
\hline
$n_f$ \quad & 
\multicolumn{1}{c|}{$c$} & 
\multicolumn{1}{c|}{$\tilde{\rho}_2$} & 
\multicolumn{1}{c|}{$\tilde{\rho}_3$} \\
\hline
$0$  & 2.31818   &   -7.066723 & -184.37823 \\
$1$  & 2.16129   &   -8.397865 & -147.27522 \\
$2$  & 1.98276   &   -9.842342 & -113.85683 \\
$3$  & 1.77778   &  -11.417129 &  -83.83139 \\
$4$  & 1.54      &  -13.144635 &  -56.87785 \\
$5$  & 1.26087   &  -15.055062 &  -32.63303 \\
$6$  & 0.928571  &  -17.190118 &  -10.67155 \\
$7$  & 0.526316  &  -19.609073 &    9.52688 \\
$8$  & 0.029412  &  -22.399086 &   28.63336 \\
$9$  & -0.6      &  -25.693806 &   47.57023 \\
$10$ & -1.42308  &  -29.709122 &   67.70445 \\
$11$ & -2.54545  &  -34.817937 &   91.25169 \\
$12$ & -4.16667  &  -41.724622 &  122.21944 \\
$13$ & -6.71429  &  -51.938541 &  168.96670 \\
$14$ & -11.3     &  -69.384046 &  252.90695 \\
$15$ & -22.0     & -108.450422 &  452.02327 \\
$16$ & -75.5     & -298.641242 & 1466.56390 \\
\hline

\end{tabular}
\end{center}
\caption{ {\textsl{
Values of the invariants for ${\cal R}_{e^+e^-}$(non-singlet) 
obtained from the calculations of Ref.~\cite{r3calc}, \cite{ccalc}, 
\cite{c2calc}--\cite{r2calc}.  
}}}

\label{invtable}
\end{table}

\newpage

\begin{table}[hbp]
\begin{center}
\begin{tabular}[b]{|r||c|c|c||c||c|}
\hline
$n_f$ \quad & $a^*$ & $\adot$ & $\ap$ & ${\cal R}^*$ & $\gamma^*$ \\
\hline
$0$  &$0.158279$& ${\scriptstyle(0.1334)}$ & ${\scriptstyle(2.50)}$  
& $0.164 \pm 0.083$ & $3.28$ \\
$1$  &$0.168688$& ${\scriptstyle(0.1465)}$ & ${\scriptstyle(1.20)}$  
& $0.174 \pm 0.083$ & $3.20$ \\
$2$  &$0.180844$& ${\scriptstyle(0.1633)}$ & ${\scriptstyle(0.832)}$ 
& $0.185 \pm 0.080$ & $3.09$ \\
$3$  &$0.195462$& ${\scriptstyle(0.1857)}$ & ${\scriptstyle(0.651)}$ 
& $0.199 \pm 0.073$ & $2.94$ \\
$4$  &$0.213910$& ${\scriptstyle(0.2162)}$ & ${\scriptstyle(0.540)}$ 
& $0.214 \pm 0.059$ & $2.73$ \\
$5$  &$0.239369$& ${\scriptstyle(0.2588)}$ & ${\scriptstyle(0.462)}$ 
& $0.235 \pm 0.028$ & $2.40$ \\
$6$  &$0.282493$& ${\scriptstyle(0.3164)}$ & ${\scriptstyle(0.402)}$ 
& $0.266 \pm 0.051$ & $1.76$ \\
$7$  &\quad---&  $0.383293$ & $0.3525$ &  $0.35 \pm 0.37$ &  \,\,\, $0$ \\
$8$  &\quad---&  $0.432267$ & $0.3094$ &  $0.48 \pm 0.64$ &  \,\,\, $0$ \\
$9$  &\quad---&  $0.429519$ & $0.2702$ &  $0.52 \pm 0.75$ &  \,\,\, $0$ \\
$10$ &\quad---&  $0.376034$ & $0.2341$ &  $0.44 \pm 0.61$ &  \,\,\, $0$ \\
$11$ &\quad---&  $0.301883$ & $0.2001$ &  $0.32 \pm 0.38$ &  \,\,\, $0$ \\
$12$ &\quad---&  $0.229746$ & $0.1673$ &  $0.21 \pm 0.21$ &  \,\,\, $0$ \\
$13$ &\quad---&  $0.166832$ & $0.1346$ &  $0.14 \pm 0.11$ &  \,\,\, $0$ \\
$14$ &\quad---&  $0.112784$ & $0.1007$ &  $0.08 \pm 0.05$ &  \,\,\, $0$ \\
$15$ & ${\scriptstyle(0.0674)}$ &  
                 $0.065248$ & $0.0642$ &  $0.043 \pm 0.015$ &  \,\,\, $0$ \\
$16$ &$0.020058$ & ${\scriptstyle(0.0215)}$ & ${\scriptstyle(0.0228)}$ 
&$0.013 \pm 0.001$ & $0.001$ \\
\hline

\end{tabular}
\end{center}
\caption{ {\textsl{
Infrared-limit results for ${\cal R}_{e^+e^-}$(non-singlet) in OPT at 
fourth order for different $n_f$ values.  For $n_f=0,\ldots,6$ and $n_f=16$ 
the limit is governed by a fixed point at $a^*$:  For $n_f=7,\ldots,15$ 
it arises from the pinch mechanism, with an ``unfixed point'' at $\adot$ and 
a ``pinch point'' at $\ap$.  
(The $\adot$ equation has solutions outside this range, giving the 
values in parentheses, but these violate the $\adot > \ap$ requirement.  Also, 
the fixed-point equation has a solution for $n_f=15$, but one that 
violates the $\gamma^* \ge 0$ requirement.)  
The last column gives values for the critical exponents $\gamma^*$ which 
characterize the power-law approach of ${\cal R}$ to its fixed-point limit; 
${\cal R}^* - {\cal R} \propto Q^{\gamma^*}$.  In the unfixed-point case one 
finds instead ${\cal R}^\star - {\cal R} \propto 1/\mid\! \ln Q \!\mid^2$.  }}}

\label{IRtable}
\end{table}

\newpage


\begin{table}[hbp]
\begin{center}
\begin{tabular}[b]{|r||d{6}|d{6}|d{6}||d{6}|d{6}|d{6}|}
\hline
$n_f$ \quad & 
\multicolumn{1}{c|} {$c a$} & 
\multicolumn{1}{c|} {$c_2 a^2$} & 
\multicolumn{1}{c||}{$c_3 a^3$} & 
\multicolumn{1}{c|} {$r_1 a$} & 
\multicolumn{1}{c|} {$r_2 a^2$} & 
\multicolumn{1}{c|} {$r_3 a^3$} \\
\hline
$0$  & 0.36692 & 0.032328 & -1.39925 & -0.25 & -0.238596 & 0.524718 \\
$1$  & 0.364583 & -0.060272 & -1.30431 & -0.25 & -0.20734 & 0.489117 \\
$2$  & 0.35857 & -0.183906 & -1.17466 & -0.25 & -0.165126 & 0.440499 \\
$3$  & 0.347489 & -0.353982 & -0.993507 & -0.25 & -0.106587 & 0.372565 \\
$4$  & 0.329421 & -0.599622 & -0.729799 & -0.25 & -0.021696 & 0.273674 \\
$5$  & 0.301813 & -0.987899 & -0.313913 & -0.25 & 0.112331 & 0.117717 \\
$6$  & 0.262315 & -1.74675 & 0.484438 & -0.25 & 0.371865 & -0.181664 \\
$7$  & 0.201733 & -3.98554 & 2.80955 & -0.228168 & 1.11073 & -0.968966 \\
$8$  & 0.012714 & -5.89311 & 5.48146 & -0.150668 & 1.72852 & -1.47106 \\
$9$  & -0.257711 & -6.75979 & 7.37992 & -0.102638 & 2.05663 & -1.74115 \\
$10$ & -0.535126 & -6.02116 & 6.90794 & -0.098434 & 1.8826 & -1.61273 \\
$11$ & -0.768429 & -4.50798 & 5.11621 & -0.122888 & 1.44444 & -1.27189 \\
$12$ & -0.957273 & -3.02694 & 3.3723 & -0.156741 & 0.999204 & -0.921032 \\
$13$ & -1.12016 & -1.83192 & 2.08731 & -0.190139 & 0.635461 & -0.631526 \\
$14$ & -1.27446 & -0.907359 & 1.20999 & -0.220064 & 0.353673 & -0.405223 \\
$15$ & -1.43545 & -0.183663 & 0.619584 & -0.245711 & 0.135041 & -0.228425 \\
$16$ & -1.51438 & 0.352822 & 0.161556 & -0.25 & -0.031878 & -0.060584 \\
\hline

\end{tabular}
\end{center}
\caption{{\textsl{
Terms in the optimized $\beta$-function and ${\cal R}$ series in the 
infrared limit ($a=\astar$ or $\adot$, as appropriate).
 }}} 

\label{coeffstable}
\end{table}

\newpage

\begin{figure}[htb]


\centering
\includegraphics[width=0.64 \textwidth]{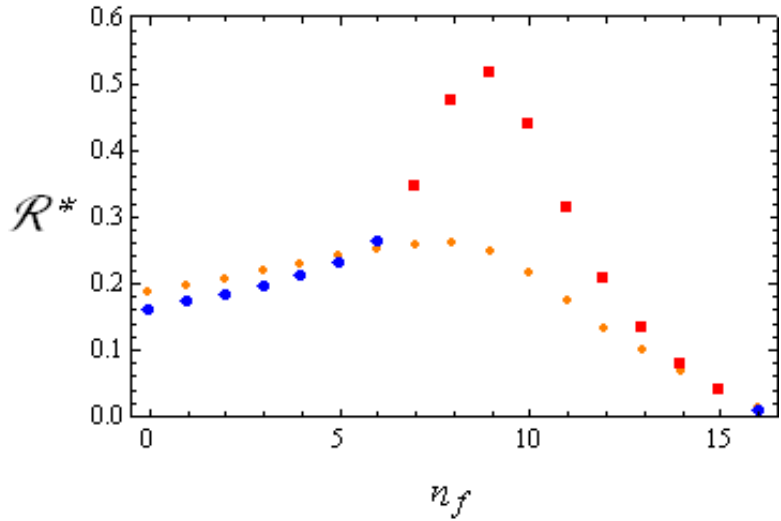}
\caption{{\textsl{
Infrared limiting values of ${\cal R}_{e^+e^-}$(non-singlet) 
in fourth-order OPT, as a function of $n_f$.  
The large dark circles and squares are, respectively, from fixed-point and 
pinch mechanisms.  (The smaller points are the corresponding results 
in the FAC scheme.)  Estimated uncertainties are large --- about 50\% at 
low $n_f$, rising to 150\% around $n_f=9$, then shrinking to about 10\% 
at $n_f=16$.   \label{Rresults}
}} }
\end{figure}

     Fig.~\ref{Rresults} plots the infrared limiting ${\cal R}$ values against 
$n_f$.  The large ``bump'' around $n_f \approx 9$ is where the pinch 
mechanism produces really dramatic spiking of ${\cal R}$ as $Q \to 0$, 
as seen in Fig.~\ref{RvsQnf8} for $n_f=8$.  If, instead of ${\cal R}^*$, 
we had plotted ${\cal R}(Q)$ for some low, but finite $Q$ --- say around 
$\half \tilde{\Lambda}_{\cal R}$ --- the bump would not have appeared and 
the points would have been close to the smaller, fainter points.  

  Those smaller points are the infrared-limiting results in the FAC (fastest 
apparent convergence) or ``effective charge'' \cite{Grunberg} scheme.  
That scheme is defined such that all the $r_m$ coefficients vanish, giving 
${\cal R} = a_{{\rm FAC}}(1+0+0+\ldots)$.  The FAC $\beta$ function's 
coefficients then coincide with the $\tilde{\rho}_n$ invariants (and so can 
be read off from Table~\ref{invtable}).  Since those coefficients do not 
evolve with $Q$, the infrared limit in FAC is simply obtained by finding 
the fixed point of the FAC $\beta$ function.  Many authors (e.g. \cite{KS}) 
have observed that, at low orders, FAC seems to yield very similar results 
to OPT.  That observation holds true here, certainly at low $n_f$ and close 
to $n_f=16\half$.  It also holds in the range $7 \lesssim n_f \lesssim 13$ 
at energies $Q \gtrsim \half \tilde{\Lambda}_{\cal R}$, as noted above.  
However, the FAC scheme does not see the extreme spiking at $Q=0$.  
While it is still true, because the error estimates (see Table~\ref{IRtable}) 
are so large in this region, that OPT and FAC infrared results agree 
within the error estimate, it is fair to say that the presence or absence of 
the spike is a qualitative difference in the predictions of the two schemes.  
(We expect other distinct differences between OPT and FAC to emerge at higher 
orders since the FAC $\beta$ function is almost certainly factorially 
divergent, whereas an ``induced convergence'' mechanism is conjectured to 
operate in OPT \cite{optult}.)

\section{Approach to the $Q=0$ limit}
\setcounter{equation}{0}

     Proper analysis of the subleading terms governing the approach to the 
$Q=0$ limit, in both the fixed- and unfixed-point cases, is surprisingly 
subtle and intricate.  We postpone details to a future publication and report 
here only the main results.  

     The usual lore is that the approach to a fixed point is described by 
a power law with a critical exponent given by the slope of the $\beta$ 
function at the fixed point:
\BE
{\cal R}^* - {\cal R} \propto Q^{\gamma^*} \quad {\mbox{{\rm with}}} \quad 
\gamma^* =\beta'(\astar) \equiv 
\left.  \frac{d \beta(a)}{d a}\right|_{a=\astar}.
\EE
The derivation of this result, in a fixed RS, and the proof that 
$\beta'(\astar)$ is invariant under RS changes \cite{Gross} is subject to some 
caveats --- which, as Ch\'yla \cite{chylafp} has rightly pointed out, are not 
necessarily to be viewed as very rare exceptions.\footnote{
Regarding some other comments in Ch\'yla's paper, note that it 
was written before the correct result for $r_2^\smmsbar$ \cite{r2calc} was 
published.  An earlier, incorrect result had made it seem that 
$\tilde{\rho}_2$ was large and positive, so that third-order OPT apparently 
failed to yield a finite infrared limit, unlike the actual situation 
\cite{CKL,lowen}. 
}
In OPT it is far from obvious that the above result will hold because the 
optimized couplant and optimized $r_m$ and $c_j$ coefficients have 
$\epsilon \ln \epsilon$ corrections as they approach their fixed-point 
limits, where $\epsilon \equiv B(a)$.  Remarkably, though, the 
$\epsilon \ln \epsilon$ terms cancel in ${\cal R}$, leaving 
\BE
{\cal R}^* - {\cal R} = \frac{\astar}{k-1} \epsilon + 
O(\epsilon^2 \ln \epsilon)
\EE
in $\kplus$ order.  
From the int-$\beta$ equation, (\ref{intbeta}), together with the 
$\boldsymbol{\rho}_1(Q)$ definition in Eqs.~(\ref{rhodefs}), (\ref{Lameff}), 
one sees that 
\BE
\ln \epsilon = \gamma^* \ln Q + {\mbox{\rm const.}}, 
\EE
so that $\epsilon$, and hence ${\cal R}^*-{\cal R}$, is proportional to 
$Q^{\gamma^*}$ where $\gamma^*$ is the slope of the optimized $\beta$ 
function at its fixed point: that is, $\gamma^* = b {\astar}^2 \sigma$
with $\sigma$ given by Eq.~(\ref{sigma}), evaluated in the optimized scheme.  
At fourth order this corresponds to 
\BE
\gamma^* = b \astar (3+ 2 c \astar + {c_2}^* {\astar}^2).
\EE
The numerical $\gamma^*$ values are reported in Table~\ref{IRtable}.  Note 
that $\gamma^*$ is around $2$ or $3$ for $0 \le n_f \le 6$, so the resulting  
low-$Q$ behaviour (Fig.~\ref{RvsQnf2}) is appropriately described as 
``freezing'' of the couplant.  However, when $\gamma^*$ is very small, as 
in the $n_f=16$ case, one sees instead ``spiking'' at $Q \to 0$, though not 
quite as extreme as the logarithmic spiking produced by the pinch mechanism.  

     In the unfixed-point case we find, after a lengthy calculation, the 
simple result:
\BE
{\cal R}^\star - {\cal R} = \frac{{\adot}^4 (3+ c \ap)}{2 {\ap}^5} \delta^2 + 
O(\delta^3).
\EE
From the int-$\beta$ equation and $\boldsymbol{\rho}_1(Q)$ definition 
we obtain (as in Eq.~(\ref{pinchdel}))
\BE \delta = \frac{\pi}{b} \frac{1}{(3+c \ap)} 
\frac{1}{\mid\!\ln Q/\tilde{\Lambda}_{{\cal R}}|} \quad\quad\, 
{\mbox{\rm as}} \,\, Q \to 0.
\EE
Therefore
\BE 
{\cal R}^\star - {\cal R} = 
\frac{1}{b_{\rm ir}^2}\frac{1}{\mid\!\ln Q/\tilde{\Lambda}_{{\cal R}}|^2} + 
O \left( \frac{1}{\mid\!\ln Q\!\mid^3} \right),
\EE
where 
\BE
b_{\rm ir} \equiv \sqrt{2 \ap(3+c \ap)} 
\left(\frac{\ap}{\adot}\right)^2 \frac{b}{\pi}.
\EE

    One way to look at the result is to note that 
\BE
Q \frac{d {\cal R}}{d Q} \sim - 2 b_{\rm ir} ({\cal R}^\star -{\cal R})^{3/2}
\EE
for ${\cal R}$ close to ${\cal R}^\star$.  Thus, the ``physically defined 
$\beta$ function'' associated with ${\cal R}$ is predicted by OPT to have 
neither a simple nor a double zero, but something in between.  An even more 
intriguing interpretation is to see the low-energy prediction as 
\BE 
{\cal R} = {\cal R}^\star - \lambda^2 
\left( 1+ O(\lambda) \right)
\EE
with $\lambda \sim 1/(b_{\rm ir} \ln Q)$ viewed as the running 
coupling constant of some infrared effective theory whose $\beta$ function 
starts $b_{\rm ir} \lambda^2 (1 + \ldots)$.

\section{Discussion}
\setcounter{equation}{0}

    We now briefly discuss the implications of our results. 
The abrupt change between $n_f=6$ and $n_f=7$ seems indicative of a phase 
transition.  For $n_f \le 6$ the phase is presumably the one we are familiar 
with in the real world; colour is confined and chiral symmetry is broken, 
with the associated goldstone bosons (pions) being massless.  Vector mesons 
($\rho$'s, etc.) have masses of order $\tilde{\Lambda}$ and their resonant 
contribution dominates $e^+e^- \to$ {\it hadrons} at low energies.  Although 
the actual $R_{e^+e^-} \propto 1+{\cal R}$ is very different from the smooth 
perturbative prediction (Fig.~{\ref{RvsQnf2}), the two agree well after 
Poggio-Quinn-Weinberg (PQW) smearing \cite{PQW} is applied to both 
\cite{lowen}.  

    For $n_f>7$ the effective low-energy theory seems to be a 
{\it renormalizable} theory with a mass scale appearing only 
in logarithms.  The extreme spiking of ${\cal R}$ as $Q \to 0$ 
(Fig.~\ref{RvsQnf8}), if viewed as a resonant peak in the vector channel, 
hints that massless vector bosons are now present.  These might be the 
gluons of an unconfined phase, or they might be massless, colourless vector 
mesons of a confined phase, perhaps with unbroken chiral symmetry.  

    Between 15 and 16 flavours our OPT results switch back from  
unfixed- to fixed-point behaviour.  However, it is much less clear that this 
indicates a phase transition.  There is hardly any qualitative difference 
between the extreme (logarithmic) spiking of the unfixed-point case and the 
very strong (fractional power-law) behaviour of a fixed-point with a very 
small $\gamma^*$.   Note that the theory with $16$ flavours (or $16.4999$, 
for that matter) is not {\it exactly} scale and conformal invariant.   While 
there is a huge range of $Q$ over which the couplant is nearly constant 
(at a value about $0.78$ of its infrared limit \cite{BZlett}), it does fall 
to zero (very slowly) as $Q \to \infty$ and it does rise (very abruptly) as 
$Q \to 0$.  

    It is beyond the scope of this paper to attempt a detailed comparison 
with the literature, but we do see some points of resemblance with other 
approaches \cite{Appelquist,Shuryak,Miransky} and with some firmly 
established results in supersymmetric QCD \cite{Seiberg,Shuryak}.  
There is also a large literature on lattice Monte-Carlo studies of QCD at 
large $n_f$ values (for recent work, see \cite{MC}).

    A quick look back at third-order OPT results is in order.  There the 
results for ${\cal R}^*$ decreased roughly linearly from $0.4$ to $0$ 
as $n_f$ increased from $0$ to $16\half$ (see Fig.~1 of Ref.~\cite{BZlett}).  
Since the uncertainties were large ($\sim 100\%$ at low $n_f$), sizeable 
changes at fourth order were not unexpected.  Nevertheless, it is an 
interesting surprise to find qualitatively different features --- particularly 
the spiking phenomenon produced by the pinch mechanism, responsible for the 
prominent bump around $n_f \approx 9$ in Fig.~\ref{Rresults}.  Previously, 
the good agreement of the third-order results with the leading $16\half-n_f$ 
(BZ) expansion led us to suggest \cite{BZlett} that that expansion might 
remain good down to very low $n_f$.  That suggestion no longer seems tenable.  
We would now expect the BZ expansion to break down around $n_f \sim 9$, 
if not sooner.  

    The fact that the fourth-order results show a {\it rise} of ${\cal R}^*$ 
with $n_f$ at low $n_f$ is interesting.  At third order the OPT fixed-point 
equations would give ${\cal R}^* \sim 2.19/b$ in the $n_f \to -\infty$ 
($b \to \infty$) limit, but that limiting form only applies for 
$n_f \lesssim -200$.   At fourth order, the large-$b$ limit of 
Eqs.~(\ref{astar3}), (\ref{Rstar3}) gives ${\cal R}^* \sim 0.84/b$, a 
formula that roughly describes the OPT results up to $n_f \approx 6$.  
(Unfortunately, the large-$b$ resummation method is fraught 
with subtleties in the infrared region and it remains unclear what it 
predicts for ${\cal R}^*$ \cite{LTM}.)   

     In closing we would like to stress that the results in this paper 
are directly the result of applying the method of Ref.~\cite{OPT} to 
the Feynman-diagram results for $R_{e^+e^-}$.  We have invented nothing new, 
nor tweaked the method in any way.  The freezing or spiking, depending 
on $n_f$, is just what {\it happens} when one solves the optimization 
equations \cite{OPT} at ever lower $Q$ values.  Achieving a finite infrared 
limit was no part of the motivation for OPT (and was never considered in 
Ref.~\cite{OPT}), so the fact that it happens is a genuine prediction --- 
and a non-trivial one, as history shows.\footnote{
see the previous footnote}
The ``pinch mechanism'' (Fig.~\ref{Bfignf8}) is another remarkable 
consequence of OPT\@.  It has serious implications beyond perturbation theory, 
because it suggests that the phase structure of QCD may not be understandable 
in the traditional language of fixed points of ``{\it the} $\beta$ function.''


\clearpage

\section*{Appendix A:  Pinch mechanism at third order}
\renewcommand{\theequation}{A.\arabic{equation}}
\setcounter{equation}{0}

     The pinch mechanism can actually occur even at third order, under 
certain restrictive conditions.  (These conditions are never satisfied in 
the $e^+e^-$ QCD case, but for other physical quantities, or other gauge 
theories, the possibility could arise.)  At third order the 
$B(a) \equiv 1+ c a + c_2 a^2$ function can obviously be re-written in 
the form 
\BE
\label{Bformk2}
B(a) = \eta \left( (a-\ap)^2 + \delta^2 \right),
\EE
with 
\BE
\label{A2}
\eta= c_2, \quad\quad\quad -2 \ap \eta = c, \quad\quad\quad 
\eta (\ap^2 + \delta^2) =1.
\EE
If $\eta=c_2$ is positive and $c$ is negative, $B(a)$ has a minimum at a 
positive $\ap=-c/(2 c_2)$ that can become a pinch point if the evolution 
of the optimized $c_2$ coefficient results in $\delta$ tending to zero as 
$Q \to 0$.  The discussion around Eqs.~(\ref{Bform}) -- (\ref{Ijform}) 
then applies, predicting the $\delta \to 0$ forms of the $I_j(a)$ 
integrals, and hence of the $B_j(a)$ functions.  At this order ($k=2$) 
Eq.~(\ref{Hdef}) yields
\BE
H_1=B-c a B_2, \quad \, H_2=B_2,
\EE
so that, substituting in Eq.~(\ref{formula}), one finds
\BE
\label{riA}
2 r_1^\star = - \frac{1+c \adot}{\adot}, \quad \, 
3 r_2^\star = \frac{c}{\adot}.
\EE
From Eq.~(\ref{A2}) with $\delta \to 0$ one obtains
\BE
\ap= -\frac{2}{c}, \quad \, c_2^\star = \frac{c^2}{4}.
\EE
Substituting in the definition of $\tilde{\rho}_2$ yields a quadratic 
equation for $\adot$:
\BE
\label{adotA}
1- \frac{4}{3} c \adot + 
4 \left( \tilde{\rho}_2 - \frac{c^2}{2} \right) {\adot}^2 =0.
\EE
The infrared limit of ${\cal R}$ can be written, using Eq.~(\ref{riA}), 
as
\BE
{\cal R}^\star = \frac{1}{6} \adot (3 - c \adot).
\EE

      As noted above, the pinch mechanism requires $c$ to be negative,   
so Eq.~(\ref{adotA}) will only have a positive root if 
$\tilde{\rho}_2 - \frac{c^2}{2}$ is negative.  Finally, the pinch mechanism 
requires $\adot > \ap$ which requires $\tilde{\rho}_2/c^2 > 13/48$ (and for 
smaller $\tilde{\rho}_2$'s the fixed-point mechanism takes over).  In 
summary, the pinch mechanism can operate at third order if and only if
\BE 
c< 0 \quad \,  {\mbox{\rm and}} \quad \,
\frac{13}{48} < \frac{\tilde{\rho}_2}{c^2} < \frac{1}{2}.
\EE


\begin{thebibliography} {99}

\bibitem{tH}
  G. 't Hooft, in {\it Deeper Pathways in High-Energy Physics}, proceedings 
  of Orbis Scientiae 1977, Coral Gables, edited by A.~Perlmutter and 
  L.~F.~Scott (Plenum, New York, 1977).

\bibitem{KSS} 
  J. Kubo, S. Sakakibara, and P. M. Stevenson, Phys. Rev. D {\bf 29}, 1682 
  (1984).

\bibitem{chylafp}
  J. Ch\'{y}la, Phys. Rev. D {\bf 38}, 3845 (1988).

\bibitem{GMLSP}
  E. C. G. Stueckelberg and A. Peterman, Helv. Phys. Acta {\bf 26}, 449 (1953);
  M. Gell Mann and F. Low, Phys. Rev. {\bf 95}, 1300 (1954).

\bibitem{OPT} 
  P. M. Stevenson, Phys. Rev. D {\bf 23}, 2916 (1981).

\bibitem{CKL}
  J. Ch\'yla, A. Kataev, and S. A. Larin, Phys. Lett. B {\bf 267}, 269 (1991).

\bibitem{lowen}  
  A. C. Mattingly and P. M. Stevenson, Phys. Rev. Lett. {\bf 69}, 1320 (1992); 
  Phys. Rev. D {\bf 49}, 437 (1994).

\bibitem{r3calc}
  P. A. Baikov, K. G. Chetyrkin, J. H. K\"uhn, and J. Rittinger, 
  Phys. Lett. B {\bf 714}, 62 (2012);  P. A. Baikov, K. G. Chetyrkin, and 
  J. H. K\"uhn, Phys. Rev. Lett. {\bf 101}, 012002 (2008).

\bibitem{OPTnew}
  P. M. Stevenson, Nucl. Phys. B {\bf 868}, 38 (2013).

\bibitem{bloom}
   E. D. Bloom and F. J. Gilman, Phys. Rev. D {\bf 4}, 2901 (1971).

\bibitem{PQW}
  E. C. Poggio, H. R. Quinn, and S. Weinberg, Phys. Rev. D {\bf 13}, 1958 
  (1976).

\bibitem{Crewther}
  R. J. Crewther and L. C. Tunstall, arXiv:1203.1321 [hep-ph].

\bibitem{BZ}
  T. Banks and A. Zaks, Nucl. Phys. B {\bf 196}, 189 (1982).

\bibitem{largeb}
  A. Palanques-Mestre and P.~Pascual, Commun. Math. Phys. {\bf 95}, 
  277 (1984);
  M. Beneke, Nucl. Phys. B {\bf 405}, 424 (1993);
  D.~J. Broadhurst, Z. Phys. C {\bf 58}, 339 (1993).

\bibitem{LTM}
  C. N. Lovett-Turner and C. J. Maxwell, Nucl. Phys. B {\bf 432}, 147 
  (1994); {\it ibid} B {\bf 452}, 188 (1995);
  P. M. Brooks and C. J. Maxwell, Phys. Rev. D {\bf 74}, 065012 (2006).

\bibitem{Appelquist}
  T. Appelquist, J. Terning, and L.~C.~R. Wijewardhana, Phys. Rev. Lett. 
  {\bf 77}, 1214 (1996).

\bibitem{Shuryak}
  E. Shuryak, Summary talk at RHIC Summer Studies, Brookhaven, July 1996; 
  arXiv: hep-ph/9609249.

\bibitem{Miransky}
  V. A. Miransky and K. Yamawaki, Phys. Rev. D {\bf 55}, 5051 (1997) 
  [Erratum-{\it ibid.} D {\bf 56} 3768 (1997)].

\bibitem{Seiberg}
  N. Seiberg, Phys. Rev. D {\bf 49}, 6857 (1994).

\bibitem{MC}
  K. Yamawaki, hep-ph 1305.6352; 
  Y. Aoki {\it et al.} (LatKMI collab.) hep-lat/1305.6006;  
  Xiao-Yong Jin and R.~D. Mawhinney, hep-lat/1304.0312;
  A. Deuzeman, M.~P. Lombardo, K. Miura, T.~N. da Silva, and E. Pallante, 
  hep-lat/1304.3245.

\bibitem{bcalc}
  H. D. Politzer, Phys. Rev. Lett. {\bf 30}, 1346 (1973);  
  D. J. Gross and F. Wilczek, {\it ibid.} {\bf 30}, 1343 (1973); 
  G. 't Hooft, report at the Marseille Conference Yang-Mills Fields, 1972.

\bibitem{ccalc}
  W. Caswell, Phys. Rev. Lett. {\bf 33}, 244 (1974);
  D. R. T. Jones, Nucl. Phys. B {\bf 75}, 531 (1974);  
  E. S. Egorian and O. V. Tarasov, Theor. Mat. Fiz. {\bf 41}, 26 (1979).

\bibitem{CG}
  W. Celmaster and R. J. Gonsalves, Phys. Rev. D {\bf 20}, 1420 (1979). 

\bibitem{c2calc}
   O. V. Tarasov, A. A. Vladimirov, and A. Yu. Zharkov, Phys. Lett. B 
  {\bf 93}, 429 (1980);
  S. A. Larin and J. A. M. Vermaseren, Phys. Lett. B {\bf 303}, 334 (1993).

\bibitem{c3calc}
  T. van Ritbergen, J. A. M. Vermaseren, and S. A. Larin, Phys. Lett. B 
  {\bf 400}, 379 (1997).

\bibitem{r1calc}
  K. G. Chetyrkin, A. L. Kataev, and F. V. Tkachov, Phys. Lett. B {\bf 85},
  277 (1979);  M. Dine and J. Sapirstein, Phys. Rev. Lett. {\bf 43}, 668 
  (1979); W. Celmaster and R. J. Gonsalves, Phys. Rev. D {\bf 21}, 3112 (1980).

\bibitem{r2calc} 
  L. R. Surguladze and M. A. Samuel, Phys. Rev. Lett. {\bf 66},
  560 (1991); S. G. Gorishny, A. L. Kataev, and S. A. Larin, Phys. Lett.
  B {\bf 259}, 144 (1991).

\bibitem{Grunberg}
  G. Grunberg, Phys. Rev. D {\bf 29}, 2315 (1984); A. Dhar and V. Gupta, 
  Phys. Rev. D {\bf 29}, 2822 (1984); C.~J. Maxwell, arXiv:hep-ph/9908463.

\bibitem{KS}
  J. Kubo and S. Sakakibara, Phys. Rev. D {\bf 26}, 3656 (1982).

\bibitem{optult} 
  P. M. Stevenson, Nucl. Phys. B {\bf 231}, 65 (1984);
  K. Van Acoleyen and H. Verschelde, Phys. Rev. D {\bf 69} 125006 (2004).

\bibitem{Gross}
  D. J. Gross, in {\it Methods in Field Theory}, edited by R.~Balian and 
  J.~Zinn-Justin (North-Holland, Amsterdam, 1976).

\bibitem{BZlett}
  P. M. Stevenson, Phys. Lett. B {\bf 331}, 187 (1994); S. Caveny and 
  P.~M.~Stevenson, arXiv: hep-ph/9705319.

\end{thebibliography}
\end{document}